\documentclass[preprint]{aastex}

\shortauthors{Mart\'inez et al.}

\usepackage{graphicx}
\usepackage{txfonts}

\begin{document}

\title{AGN population in Hickson's Compact Groups. I. Data and
Nuclear Activity Classification}

\author{M. A. Mart\'{\i}nez\altaffilmark{1} and A. Del Olmo\altaffilmark{1}}
\affil{Instituto de Astrof\'{\i}sica de Andaluc\'{\i}a, CSIC, Apdo. 3004,
18080 Granada, Spain}
\email{geli@iaa.es, chony@iaa.es}
\author{R. Coziol\altaffilmark{2} }
\affil{Departamento de Astronom\'{\i}a, Universidad de Guanajuato,
Apdo. 144, 36000 Guanajuato, Mexico}
\email{rcoziol@astro.ugto.mx}
\and
\author{J. Perea\altaffilmark{1}}
\affil{Instituto de Astrof\'{\i}sica de Andaluc\'{\i}a, CSIC, Apdo. 3004,
18080 Granada, Spain}
\email{jaime@iaa.es}

\begin{abstract}

We have conducted a new spectroscopic survey to characterize the
nature of nuclear activity in Hickson Compact Groups (HCGs) galaxies
and establish its frequency. We have obtained new intermediate
resolution optical spectroscopy for 200 member-galaxies and corrected
for underlying stellar population contamination using galaxy
templates. Spectra for 11 additional galaxies have been acquired from
the ESO and 6dF public archives and emission line ratios have been
taken from the literature for 59 galaxies more.  Here we present the
results of our classification of the nuclear activity for 270
member-galaxies, which belong to a well defined sample of 64 HCGs. We
found a large fraction of galaxies, 63\%, with emission lines.  Using
standard diagnostic diagrams, 45\% of the emission line galaxies were
classified as pure AGNs, 23\% as Transition Objects (TOs) and 32\% as
Star Forming Nuclei (SFNs). In the HCGs, the AGN activity appears as
the most frequent activity type. Adopting the interpretation that in
TOs a Low Luminosity AGN coexists with circumnuclear star formation,
the fraction of galaxies with an AGN could rise to 42\% of the whole
sample. The low frequency (20\%) of SFNs confirms that there is no
star formation enhancement in HCGs. After extinction correction we
found a median AGN H$\alpha$ luminosity of 7.1$\times$10$^{39}$ erg
s$^{-1}$, which implies that AGNs in HCG have a characteristically low
luminosity. This result added to the fact, that there is an almost
complete absence of Broad Line AGNs in Compact Groups (CGs) as found
by \citet{mar08a} and corroborated in this study for HCGs, is
consistent with very few gas left in these galaxies. In general,
therefore, what may characterize the level of activity in CGs is a
severe deficiency of gas.
\end{abstract}


%
\section{Introduction}
%

From observations and theoretical simulations it is recognized that
gravitational interactions between galaxies must play a fundamental
role in the evolution of galaxies. Through star formation,
galaxy-galaxy interactions and mergers can influence the mass assembly
of galaxies and generate morphological transformations.  These
processes may also play some role in determining the type of AGN
(Active Galactic Nuclei) activity.

The connection between gravitational interactions and enhancement in
star formation in galaxies is already well documented
\citep{ken87,kee93,bar00,woo07,lon84,lin07}. However not all
galaxy-galaxy interactions are equally efficient in producing an
enhancement of star formation \citep{berg03}. Simulations \citep[][and
references therein]{mat08} have shown that this depends on various
parameters describing the initial conditions of the interactions, like
the disk stability of the interacting galaxies, the total mass or
relative mass of the interacting components \citep{cox08}, the gas
content of the galaxies or their relative orbit parameters
\citep{mat07}.

Many authors have also found evidence suggesting a connection between
circumnuclear starburts and AGN \citep[see][for review, and references
therein]{sto08}. This connection is supported by theoretical
considerations. Simulations of interacting galaxies and mergers
\citep{byr87,spr05,mat05} suggest that apart from a central burst of
star formation the inflow of gas towards the nucleus of the galaxies
could have a positive feed back on a preexisting black hole,
increasing its activity. Simulations have shown also that tidal
torques, produced during slow galaxy encounters, can transport gas
efficiently towards the center of the galaxies
\citep{mih96,spr05,mat07}.  However, it is still not clear what
fraction of the gas will form stars, increasing the bulges of the
galaxies, and what fraction will go deeper to the center of the galaxy
to nourish a black hole. Neither it is clear if these processes are
really simultaneous or competitive. In particular, we do not know yet
if there are physical conditions or galaxy environments under which
AGNs may be favored over star formation.

Early works on AGNs suggested that Seyfert activity, without
discriminating between Seyfert 1 and Seyfert 2 (hereafter Sy1 and
Sy2), appears more frequently in optically disturbed and interacting
systems \citep{ada77,dah85,kee85}. The results of more recent studies
seem less clear on this issue. Some authors
\citep[i.e.][]{mol95,lau95,dom05,woo07,kuo08} find excesses of AGNs
in interacting galaxies, whereas others claim the opposite
\citep{bus87,fue88,sch01,ho03,mil03}. Different studies propose that
the discrepancy can be explained taking into account the scale of the
environment, discriminating between small ($\sim$ 100 kpcs) and large
scales (typically until $\sim$ 1 Mpc), and the type of nuclear
activity \citep{dul99,kou06,sor06,gon08}. On large scales, they found
that no difference between the environments of the different AGN types
can be seen. On small scales, however, they found that there is a
statistically significant higher fraction of Sy2 than Sy1 with close
companions.  On this matter, we recently found that the environment
may be more important than previously thought. Indeed, we report
\citep{mar08a}, that there is a remarkable deficiency of broad-line
AGNs as compared to narrow-line AGNs in two different samples of
Compact Groups of galaxies (CGs), suggesting that tidal or group
interaction effects could influence in the AGNs in CGs. 

Recent observational studies have shown that a significant fraction of
nearby galaxies host an AGN and this fraction increases as the
luminosity of the AGN goes down
\citep{ho97,mil03,ho08,olm08,mar08b}. However, as we mention before
the relationship between galaxy interaction and nuclear activity is
open to debate. To explore further this possible connection we have
undertaken a new spectroscopic survey of galaxies in Hickson Compact
Groups \citep[][HCGs]{hic82}. The main goals of our study are: 1) to
characterize the type of nuclear activity present in the galaxy
members of HCGs and estimate more reliably its actual frequency; and
2) to establish the relation this activity may have with the
properties of the host galaxies and its level of evolution in the
group, as well as with the dynamical properties of the group itself.

The present article concentrates on the first goal. We present the
classification of nuclear activity for a sample of 270 member-galaxies
in 64 HCGs. For this study, we have obtained new long-slit
intermediate resolution optical spectroscopic observations for 200
galaxies. All these new spectra were corrected for stellar population
contamination subtracting proper galaxy templates. Spectra for eleven
additional galaxies were acquired from the ESO and 6dF public
archives. To complete our information we also used line measurements
for 59 member-galaxies more as determined in previous studies
\citep{kim95,aoki96,verd97,coz98,coz00,shi00,coz04}. A following
article will be dedicated to the analysis of the relationship between
the detected nuclear activity and the properties of the host galaxies
and their parent groups, as well as the comparison with other
environments.

The organization of this paper is the following: in Section~2 we
explain the selection of the sample; in Section~3 we present the new
spectroscopic observations and give a general description of the
reduction processes used; in Section~4 we present the results obtained
for the whole sample of 270 galaxies; the final nuclear activity
classification adopted is presented in Section~5, followed by our
conclusions in Section~6.

%
\section{Description of the sample}
%

In 1982, \citet{hic82} produced a catalogue of 100 CGs selected from
the Palomar Sky Survey. The three criteria used by Hickson to define
CGs form today the basis (with some variance) of many subsequent
studies on the same subject. The first criterion is one of population
and magnitude concordance: the groups must be composed of at least
four galaxies, with a difference, at most, of three magnitudes. The
second criterion is one of isolation: the radius of the smallest
circle containing all the galaxies in a group should be three times
smaller than the distance from the group center to the nearest galaxy
with less than 3 magnitudes in difference. The last criterion is one
of compactness, expressed in terms of surface brightness: the mean
surface brightness of the group should be brighter than 26 mag
arcsec$^{-2}$. Subsequent spectroscopic observations
\citep[]{hic92,hic97} have shown that not all galaxies in the original
HCGs have concordant redshifts. Therefore, only 92 HCGs have three o
more galaxies with concordant redshift. A detailed description of this
new whole sample as well as the possible biases introduced by the
visual search procedure and the original criteria used is reviewed in
\citet{hic97} and references therein.

\begin{figure}
\includegraphics[angle=-90,scale=0.38]{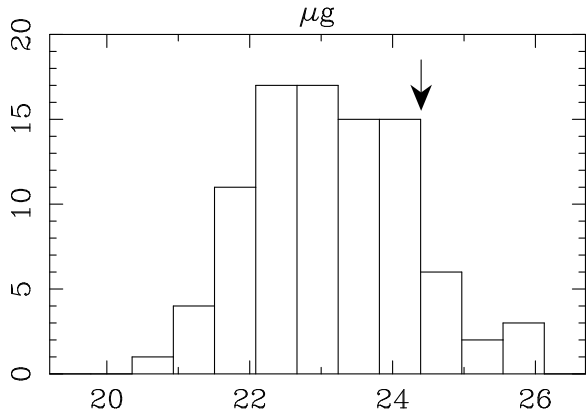}
\caption{Mean surface brightness distribution for the Hickson
catalogue. The arrow shows the
limit up to which the catalogue can be considered complete.\label{fig1}}
\end{figure}

\begin{figure}
\includegraphics[angle=-90,scale=0.38]{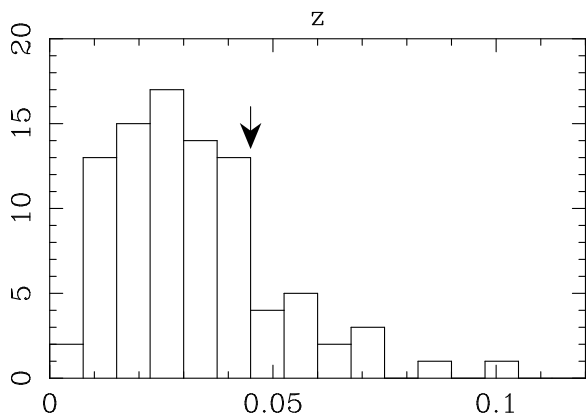}
\caption{Histogram of the redshift distribution for the Hickson
  catalogue of Compact Groups. The arrow shows the
limit up to which the catalogue can be considered complete. \label{fig2}}
\end{figure}

To carry out our analysis we have selected a well defined sample of 65
HCGs, starting with the 92 HCGs with concordant redshifts and using
two additional constraints to minimize the effects of incompleteness
in mean group surface brightness and redshift.  Hickson (1982) found
in the original sample that the catalog becomes incomplete for mean
group surface brightness fainter than about 24 mag/arcsec$^{-2}$. As
it can be seen in Figure~\ref{fig1}, where we have plotted the mean
surface brightness distribution for the whole sample (92 groups),
there is a clear deficit of low surface brightness groups and the
catalog appears incomplete for groups with about $\rm{\mu_g \ \simeq \
24.4\ mag/arcsec^2}$. Also the redshift distribution, plotted in
Figure~\ref{fig2}, shows that the catalog becomes clearly incomplete
for groups with redshifts higher than 0.045. Both limits are marked
with an arrow in each respective figure.  For our sample we have
therefore excluded all groups with $\mu_g > 24.4$ mag/arcsec$^2$
(regardless of the redshift) and those groups with z\ $> 0.045$,
obtaining a final sample of 65 HCG groups.

Based on the original \citet{hic82} catalogue, these 65 groups
contained in total 290 galaxies. However, after later revision
\citep{hic92}, only 269 member galaxies remained. For completeness
sake, we have rechecked the membership of the galaxies in our sample
using our own spectral observations. We found that H3c ($z = 0.0248$)
and H5d ($z = 0.0415$) have concordant redshifts compared to their
respective parent groups, and should not have been excluded. The same
was found to be true of H48c ($z = 0.0115$), H51f ($z = 0.0266$), H51g
($z = 0.0254$) and H72f ($z = 0.0437$).

In the case of HCG43, considered as a quintet by \citet{hic92}, we
found a discrepant redshift for H43e, namely $z = 0.0966$ compared to
$z = 0.033$ for its parent group. This makes HCG43 a quartet. We also
confirmed that H95b is a foreground galaxy, as pointed out already by
\citet{igl98}.

Some new members have also been added. We have included NGC1208, with
a projected distance of 270 kpc and a $\rm{\Delta v_r = 450 km/s}$
with respect to the group velocity, as a member of HCG23. In HCG31, we
confirm the membership of H31G as noted by \citet{rub90}, and H31Q and
H31AN505 as in \citet{verd05}, making it a sextet. \citet{bar98} found
H57h to belong to the group, which we confirm with our
spectroscopy. In our sample we have also included NGC7320c in HCG92,
as was previously suggested by \citet{mol97} and \citet{sul01}, H97-2
in HCG97, as suggested by \citet{carv97} and H100d as suggested by
\citet{hic93}. HCG21 was considered a quintet in \citet{hic82} and a 
triplet in \citet{hic92}. We have considered this group as a quintet,
because the excluded galaxies (H21d and H21e) have a difference in
radial velocity of less than 1000\ km/s with respect to that of the
group.

To summarize, after revision of the membership, we added 16 galaxies
and excluded 2 (H43e and H95b) obtaining a total of 283 galaxies.

In Table~\ref{tab1} the selected groups are listed with their main
characteristics. Columns~2 and 3 contain the position coordinates
$\alpha$ and $\delta$ (J2000) respectively. In column~4 we give the
number of galaxies considered by us as members in each group. In
column~5 we give the number of galaxies for which we have obtained
spectroscopic observations and identify those for which the spectral
information comes from the bibliography or public archives. Since no
spectrum was available for any of the galaxy member of the southern
group HCG63 (a triplet), our final sample is reduced to 280 galaxies
in 64 groups. The mean radial velocity for each group in column~6, was
recalculated taking into account the new obtained redshifts and final
member-galaxies.

The final selected sample of 64 HCGs contains 54 multiplets (84\%) and
10 triplets. A similar percentage is found for the whole HCG
catalogue. The morphological type of the galaxies were obtained from
the HyperLeda database \citep{pat03}. We find 53\% early-type (E or
S0, T$<$0), 32\% early spirals (S0a-Sbc, 0 $\ge$ T $<$ 5) and 15\%
late spirals (T $\ge$5, i.e. $\ge$ Sc). Therefore, early-type galaxies
are dominant in our sample, as in the HCG catalog. We also find that
82\% of the galaxies are brighter than M$_{\rm B} =-19$, with a median
value of $-20.15$ magnitudes. The group velocity range goes from 1387
km/s to 12966 km/s.

A value of H$_0$\ =\ 70\ km\ s$^{-1}$\ Mpc$^{-1}$ is used
throughout this article.

%
\section{Observations and Data Reduction}
%

%
\subsection{Spectroscopic Observations}
%

Two-dimensional long-slit spectroscopy were obtained for 200 galaxies
in our sample. Four telescopes were used, during various observing
runs from 2003 to 2007: 1) the 2.2m telescope in Calar-Alto
(CAHA\footnote{The Centro Astron{\'o}mico Hispano Alem\'an is operated
jointly by the Max-Planck Institut fur Astronomie and the IAA-CSIC.},
Spain); 2) the 2.56m NOT\footnote{ALFOSC is owned by the IAA and
operated at the Nordic Optical Telescope (NOT), under agreement
between IAA and the NBIfAFG of the Astronomical Observatory of
Copenhagen.} at the Observatorio del Roque de los Muchachos (ORM) in
La Palma (Spain); 3) the 2.12m telescope at the Observatorio
Astron\'omico Nacional in San Pedro M\'artir (SPM) in Baja California
(M\'exico); 4) and the 1.5m telescope\footnote{The 1.5m is operated by
the IAA in the Sierra Nevada Observatory.} at Sierra Nevada
Observatory (OSN) en el Observatorio de Sierra Nevada in Granada
(Spain).

We obtained medium resolution spectroscopy in the optical range
3600-7200 \AA, which includes the main nebular emission lines used for
standard spectroscopic classification: namely, from [OIII]$\lambda
5007$ \AA\ to [SII] $\lambda \lambda 6717, 6731$ \AA\ doublet,
including [OI]$\lambda 6300$ \AA, [NII]$\lambda 6584$ \AA\ and the
Balmer H$\beta$ and H$\alpha$ lines.

Depending on the telescope and the weather conditions, a long slit
with a width varying between 1 and 2.5 arcseconds was placed across
the nucleus of each galaxy. Except for the SPM observations, where the
slit was kept in the E-W direction, the slit was usually oriented
along the major axis of the galaxies. In some groups, however, the
slit was rotated in order to be able to observe more than one galaxy
at a time. In these cases, both galaxies were located near enough to
be observed together, but were sufficiently spatially separated in the
slit to be clearly resolved. In most cases, the galaxies also have
similar magnitudes, which allows reaching comparable signal to noise
ratio with the same exposure time.

In Table~\ref{tab2} we summarize for each observatory (column~1) and
telescope (column~2), the spectrograph (column~3) and detector used
(column~4), as well as the setup for the observation, including the
CCD characteristics, like the pixel size (column~5) and total CCD
dimension (column~6). The plate scale appears in column~7, followed by
the grating (column~8), the corresponding coverage in wavelength
(column~9), and reciprocal spectral dispersion (column~10). Whereas
only one grating was used to cover the full spectral range at the ORM,
SPM and OSN, for the CAHA observations two different gratings, B100
and G100, were necessary to get the same spectral coverage, with
similar spectral dispersion.

Exposure times were estimated considering the corresponding magnitudes
of the individual galaxies in order to reach a signal to noise ratio
(S/N) in the continuum greater than 10. The total exposure time
required was usually split in three exposures to reduce the number of
cosmic rays, and eliminate them by combining the different exposures
of the same galaxy.

In Table~\ref{tab3} we present the journal of the observations. For
each galaxy we list the date of the observation in column~2. Note that
some galaxies were observed more than once using different
telescopes. This was done to check the consistency of our results and
in some cases, when the previous observations were judge insufficient,
to repeat the observation. For each observation we identify the
grating used (column~3), the position angle (PA) of the slit
(column~4), the total exposure time (column~5), the effective airmass
(column~6) and the slit width used (column~7). In column~8, we give
the final aperture, in kpcs, adopted to extract the nuclear spectrum
from the original two dimensional data. Reduced unidimensional spectra
were used to obtain the spectral activity classification. In column~9
we have also included information about the presence or absence of
emission lines in each galaxy.

In the subsequent subsections we explain in more details the
observation made in each observatory.

%
\subsubsection{CAHA}
%

Most of the data were obtained in four runs from 2004 to 2006 with
the CAFOS spectrograph at the 2.2m of Calar Alto (CAHA), equipped
with a SITE $2048 \times 2048$ elements CCD detector having a pixel
size of $24\ \mu$m. Two grating settings were used to cover the full
spectral range. The B100 grating provides a wavelength coverage from
3200\AA\ to 5800\AA\ and the G100 grating one from 4900\AA\ to
7800\AA. Both have a reciprocal dispersion of 2 \AA/pixel and
spectral resolution of 4.8\AA. In the spatial direction along the
slit, a single pixel on the detector has a projection of 0.53'' on
the sky. A slit width of 1.5-2'' was used according to the seeing
condition, which was generally better than 1.2 arcseconds.

In all, we have observed 96 galaxies of our sample with this
telescope: 64 galaxies were observed using both configurations (B100
and G100) and 32 using only the G100 grating. Note that due to the
redshift of the galaxies, even using only the G100, it was possible, in
the majority of the cases, to measure the H$\beta$ and [OIII]$\lambda
5007$\AA\ lines necessary for spectral classification. All the spectra
obtained in CAHA have been used to classify the nuclear activity.

%
\subsubsection{ORM}
%

The observations using the 2.56m Nordic Optical Telescope (NOT) were
mainly obtained during two runs, in November 2005 and April 2006.  But
we have also included here previous observations, not yet published,
which were made in 1999 (seven galaxies in H26) and 2000 (nine
galaxies in different groups). All of our observation used the ALFOSC
spectrograph. This spectrograph is equipped with a E2V $2048
\times 2048$ CCD detector, with a pixel size of $13.5\mu$m and plate
scale of 0.19''/pixel. We used the GR4 grism, with spectral coverage
of 3200-9100\AA\ and spectral resolution of 8\AA. Some galaxies have
also been observed with the GR8 grism, which covers the H$\alpha$
region from 5800\AA\ to 8300\AA, with a spectral resolution of 3\AA.

A total of 55 galaxies of our sample were obtained with this
telescope: 53 were observed with the GR4 grism, two galaxies (H26a,
H40c) with the GR8 and one (H40d) has been observed with both grisms.

%
\subsubsection{SPM}
%

Observations with the 2.12m telescope of SPM were obtained during
three runs: two campaigns in 2004 and one in May 2005. The Boller \&
Chivens spectrograph used was equipped with a SITE $1024 \times 1024$
CCD detector, with a pixel size of $24 \mu$m and a plate scale of
1.05''/pixel. With only one grating, R300, we could cover the whole
spectral range, from 3500-7500\AA, at a spectral resolution of 11\AA.

The long slit was centered on the most luminous part of the galaxies
and aligned in the east-west direction, to minimize errors introduced
by guiding. We have observed with this telescope 57 galaxies belonging
to the sample.

%
\subsubsection{OSN}
%

We have also obtained spectra during several runs from 2005 to 2007
with the 1.5m telescope at the Observatory of Sierra Nevada in
Spain. The ALBIREO spectrograph was equipped with a Loral/Lesser $2048
\times 2048$ CCD detector with a pixel size of $15 \mu$m and a plate
scale of 0.9''/pixel. With only one grism, Red4, we could observe the
whole spectral range, from 3600\ \AA\ to 7500\ \AA\ at a spectral
resolution of 5\AA.

We have observed 65 galaxies of the sample with this configuration.
Most of them were observed previously with other telescopes, but in
some cases under bad weather conditions. Repeating the observations
were consequently considered important to confirm their spectral
classification.

%
\subsection{Reduction procedures}
%

The spectra were reduced using the two-dimensional spectrum reduction
package in IRAF\footnote{IRAF is distributed by the National Optical
Astronomy Observatories, which are operated by AURA, Inc., under
contract with the National Science Foundation}, following standard
procedures.

We have built one bias for each night for each observational campaign
by combining 10-20 biases taken each night. This bias was subtracted
from individual spectra after overscan removal. Flat-field correction
was done using a composite flat. Our flat-field reduction takes into
account the pixel-to-pixel sensitivity variations of the CCD used and
the non-homogeneous illumination effect produces by the slit (using
twilight flat-fields). After flat-field correction a median 2D
spectrum was obtained for each galaxy by combining the different
exposures, eliminating in this way most of the cosmic rays.

Wavelength calibration was done in a standard bidimensional way in
IRAF, using comparison lamp observations done at the same telescope
position as the galaxies. Flux calibration was performed using
STANDARD and SENSFUNC tasks in IRAF. Several spectrophotometric
standard stars from the lists of \citet{oke90,ham92,ham94,mas88} were
observed each night for this purpose. When the nights were photometric
we calculated a new extinction curve to obtain a better absolute flux
calibration, using 3-4 spectrophotometric standard stars with
different airmasses. The flux calibration was applied to each galaxy
on the two dimensional spectrum. After flux calibration, a two
dimensional spectrum without sky contribution was obtained using the
APALL subroutine. This step also allows correcting for spatial
curvature along the slit. However, this correction is not important in
our observations because the galaxies are located at the central
spatial sections on the slit. Once we obtained the spectrum of each
galaxy, corrected from instrumental response and calibrated in
wavelength and flux, we extracted the one-dimensional spectrum in
order to measure the flux of the emission lines needed to do the
spectral nuclear classification.

\begin{figure}
\plotone{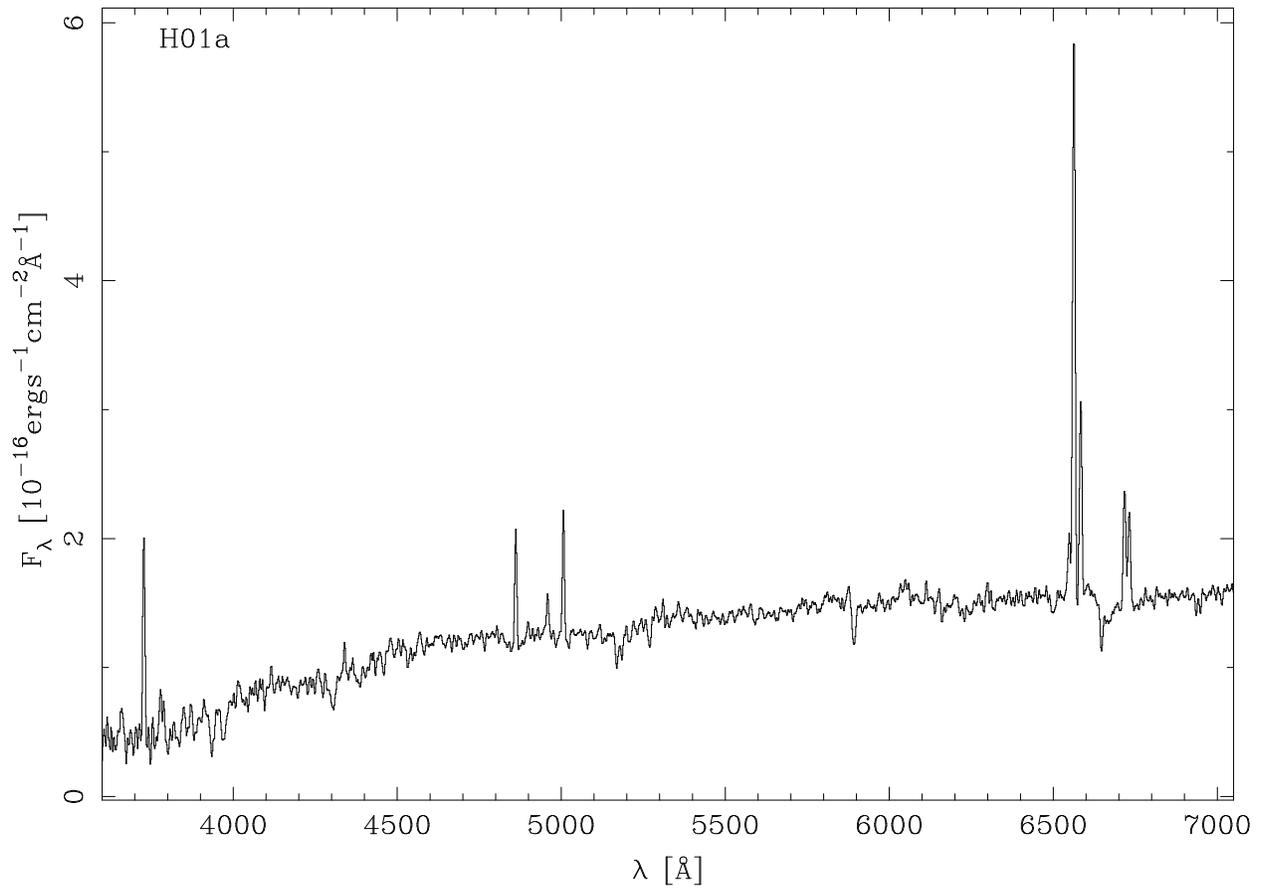}
\caption{Example of the one-dimensional spectrum corresponding to HCG1a. [See the electronic edition of the Journal for Figs. 3.1 to 3.200].\label{fig3}}
\end{figure}

The one-dimensional extracted spectra for all the observed galaxies are
represented in the Figure Set \ref{fig3}.

%
\subsection{Template subtraction and line measurement}
\label{sec-3.3}
%

Nuclear spectra of galaxies can be contaminated by absorption features
produced by stellar populations in the host galaxy, having
intermediate and older ages. When we have emission, this contamination
dilute the flux of the more important lines, affecting their ratios
and, consequently, their classification. For instance, as the
contribution of the bulge component becomes more important, absorption
features can hide the presence of weak emission lines, the effect
increasing with the aperture of the slit. Usually this contamination
should affect only the Balmer lines, but due to the broadness and
intensity of some of the absorption features and to the spectral
resolution used, it can also affect nearby emission lines, like
[NII]$\lambda\lambda$6548,6584\AA. In CGs, the majority of the
galaxies show Balmer absorption lines and correction for the
contamination by underlying stellar population is especially
important. In particular, many of the galaxies with apparently no or
very weak emission features, show, after template subtraction to
correct for the absorption dilution, evidence consistent with low
luminosity AGNs \citep[hereafter LLAGN;][] {coz98,mar08a}. Therefore,
to obtain a more solid classification of the type of activity
present in the galaxies in our sample, we have systematically applied
a template correction.  As templates, we used the spectra of galaxies
without emission which were observed at the same time with the same
setup as the emission line galaxies.

We use four different templates for each telescope configuration.
These templates show no emission lines, as verified by subtracting the
other templates, and have high S/N. Usually they correspond to
non-emission galaxies of the HCG observed sample. In some observing
runs we also used as template the non-emission galaxy NGC1023, which
was frequently used by other authors for such purpose
\citep[i.e.][]{ho97}. After inspection of all the spectra for each run
and setup, the galaxies selected as templates were: H7b, H57c, H93d
and NGC1023 for the CAHA observations; H3d, H48a, H79c and H97d for
the ORM; H61d, H97a, H98c and NGC1023 for the OSN; and H57f, H82a,
H92b and H99b for the SPM observations. The four templates are
subtracted from the spectra of each target galaxy, which yields four
different corrected spectra per galaxy.

The method to subtract the templates is described in
\citet{coz98,coz04}. All the spectra are first shifted to zero
redshift. Then, two spectral regions of interest for classification,
namely one centered on H$\beta$ line, with range 4500-5500\AA, and one
centered on H$\alpha$, with range 6350-6850\AA, are extracted.  Each
spectrum is then normalized by the corresponding fitted continuum in
the equivalent regions of each galaxy and the templates are
subtracted. Finally, we multiply each subtracted spectrum by the
previously fitted continuum to obtain the original calibrated flux
units and measure the line fluxes. Although in principle this
procedure is not necessary for those galaxies with strong emissions,
or those not showing emission lines, we have performed it to all the
spectra for consistency sake and also to verify that no weak emission
lines were missed.

After template subtraction, the emission line fluxes were measured in
an interactive way using the SPLOT routine in IRAF. The lines measured
are the following: H$\beta$, [OIII]$\lambda 5007$\AA, [OI]$\lambda
6300$\AA, [NII]$\lambda\lambda 6548, 6584$\AA, H$\alpha$ and the
doublet [SII]$\lambda\lambda 6717, 6731$\AA. Because the [OI] and the
two [SII] lines are not affected by absorption features they were
measured without template subtraction and their errors were obtained
by quadratic addition of the photo-counting errors and one sigma error
of the local continuum. All the other lines were measured after the
template subtraction. The emission lines were measured directly over
the spectrum without template subtraction for H26a and H40c, two SFN
galaxies only observed in the [NII] and H$\alpha$ region with GR8
grating, because no templates were available for this setup.

The lines were measured by fitting a Gaussian. In the case of the two
[NII] lines and H$\alpha$ line, we used the deblend option in the
SPLOT task, constraining our fit by using the same FWHM that was
obtained for the [SII] lines. When the [SII] lines were too noisy or
not available, due mainly to the presence of the atmospheric
absorption band, we used the FWHM of the [OIII]$\lambda 5007$\AA\ as a
constraint. When a single gaussian fitting did not seemed to be
sufficient, we used the NGAUSFIT procedure in IRAF to perform a
multiple fit Gaussian components. This task was also used to check for
the presence of possible broad components as explained in
\citet{mar08a}. The final flux was estimated by calculating the
median of the four obtained values after template subtraction,
adopting the average deviation as the uncertainty.

For each galaxy with emission lines, we list in Table \ref{tab4} the
emission line ratios, in log scale, that were used to determine the
nuclear classification: [OIII]/H$\beta$ (column~2), [OI]/H$\alpha$
(column~3), [NII]/H$\alpha$ (column~4) and [SII]/H$\alpha$
(column~5). In column~6 we also include the
H$\alpha$ flux. In column~7 the observed H$\alpha$
luminosity, without extinction correction, is also estimated. Finally
in column~8, we list the nuclear classification we adopted, based on
the different standard diagnostic diagrams (see Section\
\ref{sec-5.2}).

%
\section{Archive spectra and data from the literature}
%

We have inspected several spectroscopic public archives to complete
the number of available spectra of galaxies in our sample. We found
spectroscopic data for eleven more galaxies: six in the ESO public
archive (H4b, H4d, H32a, H32b, H32d, H91d) and five in the 6dF Archive
(H21a, H21b, H21d, H91a, H91b). These spectra have only relative flux
calibration. Six galaxies show emission: H4b, H4d, H21a, H21b, H91a
and H91b. For these galaxies we have measured the emission line
intensities without a template subtraction. The corresponding measured
emission line ratios and nuclear classification are given in Table
\ref{tab5}. For H32a, H32b, H32d, and H91d only the blue part of
the spectrum, within the range 4000-6000\AA, seemed available. Because
the spectra show no obvious signs of emission, these galaxies were
classified as non-emission. For the 6dF spectra, only the red part of
the spectrum, within the range 5500-8000\AA, is available. H21d was
classified as non-emission. H21a, H21b and H91b show evidence of
narrow emission lines. In H91a we also detected a broad component for
H$\alpha$ emission line. For this galaxy, the blue spectrum in the ESO
archive shows a broad H$\beta$ component. In fact, this galaxy was
already classified as a Seyfert 1.2 by \citet{dah85}. It is important
to note that this is the only galaxy in our sample which shows the
presence of genuine broad emission lines.

For 59 galaxies in our sample, we have collected the line emission
ratios from the literature. For 55 galaxies, we took the values listed
in \citet{coz98,coz04}. The line ratios for H92c and H96a were taken
from \citet{aoki96} and \citet{kim95} respectively, and for H96d they
were taken from \citet{verd97}. For H21c no published ratios seem
available in the literature. since it was observed previously by
\citet{coz00}, we have adopted their nuclear classification.

In Table~\ref{tab5} we give the line ratios for the extra
galaxies taken from the literature, together with the ones we measured
on archive spectra (as marked by an asterisk).

%
\section{Nuclear Activity Classification}
%

Adding to our own spectroscopic observations (200 galaxies) the 11
archive spectra and the emission line ratios of the 59 galaxies as
found in the literature, we obtained spectral information for 270 of
the 280 galaxies (96\%) of our HCG sample. The remaining 10
galaxies without spectra turned out to be the fainter (m$\geq$ 17)
members of their parent group (H13e, h24d, h24e, h74e, h76g, h94e,
h94g) and/or to be unobservable from the north (h21e, h32c, h91c).  No
data could be found in the literature or in archives.

Of the 270 galaxies, 63\% (169 galaxies) show emission lines: 122 are
new observations, 6 are measurements from the archive spectra and 41
come from the literature. In this section we describe the process we
followed for classifying these galaxies.

%
\subsection{Spectral classification throught Diagnostic Diagrams}
\label{sec-5.1}
%

The standard method for narrow emission-line galaxy classification is
based on the empirical diagnostic diagrams of line ratios, first
introduced by \citet{bal81} and subsequently refined by
\citet{vei87}. These last authors established different empirical
separation sequences using published data for known giant HII regions,
Sy2 and LINERs. The most useful diagrams are those that use the ratios
of very close emission lines, for example
[OIII]$\lambda$5007/H$\beta$, [NII]$\lambda$6584/H$\alpha$,
[SII]($\lambda\lambda6717+6731)$/H$\alpha$ and [OI]$\lambda$6300
/H$\alpha$. This choice makes them insensitive to reddening and errors
in flux calibration. The different empirical separation sequences
allow to distinguish between different ionization mechanisms, which
are responsible for producing the emission. In particular, they allow
to separate photo-ionization by massive OB stars, typical of Star
Formation Nuclei (SFNs), from more energetic sources like AGNs (Sy2 or
LINER). The most used diagram relates [OIII]/H$\beta$
with [NII]/H$\alpha$, and is recognized as the BPT
diagram. There exist other useful diagrams like
[OIII]/H$\beta$ versus [OI]/H$\alpha$,
hereafter identified as the [OI]-diagram, and
[OIII]/H$\beta$ versus [SII]/H$\alpha$, hereafter
identified as the [SII]-diagram.

Although diagnostic diagrams are useful in identifying the main source
of ionization in galaxies following the empirical distributions
devised by \citet{vei87}, there remains some uncertainty for objects
occupying the border regions, or for galaxies that can be classified
differently according to one or another diagnostic diagram. In
particular, \citet{ver97} and
\citet{ho97} introduced the term Transition Objects (TOs) to
characterize emission-line galaxies with line ratios intermediate
between SFNs and AGNs. They suggested that in TOs galaxies an AGN
(Seyfert or LINER) coexists with circumnuclear star formation
regions. The coexistence of star formation in the circumnuclear region
of these galaxies dilutes the AGN signature producing the TO
spectrum. This interpretation is partly supported by high spatial
resolution data. In studies of nearby circumnuclear star formation
regions, for example, some authors distinguish a transition in the
emission line ratios from the central AGN regime to the SFN one
\citep{ben06,zut07}. Although the physical origin of TOs continues to
be an open discussion \citep{shi07}, optical and radio observations
strongly suggests that TOs harbor AGNs \citep{fil04,nag05,ho08}.

Sequences defined by \citet{vei87} have been extensively used during
many years. However, more recently, some authors have defined new
separation sequences that can be used in diagnostic diagrams. In
\citet[][hereafter Ke01]{kew01} a combination of
stellar population synthesis models and detailed photo-ionization
models were used to establish theoretical maximum star formation limit
for each diagnostic diagram (BPT, [OI]-diagram and
[SII]-diagram). According to Ke01, galaxies lying above these
sequences are obviously dominated by an AGN. Later,
\citet[][hereafter Ka03]{kau03} showed that the Ke01 line is well
above the star formation sequence delineated by a sample of SDSS star
forming galaxies and used this characteristic to define a new
empirical curve for pure star forming galaxies. Any galaxies below
this curve are obviously dominated by star formation.  In
\citet{kew06}, the authors have subsequently defined a new
classification scheme (see their Fig. 1), based on the two previously
defined sequences (Ke01 and Ka03), to discriminate between SFNs, AGNs
and TOs. According to \citet{sta06}, there is effectively a region
below the Ke01 sequence which cannot be explained by star formation
models with any realistic combination of metallicity, ionization
parameters and extinction. They suggested that even below the Ka03
sequence there might be galaxies with an AGN contributing up to 3\% in
H$\beta$. Based on a large SDSS sample and using a grid of
photo-ionization models, \citet{sta06} asserted that the BPT diagram
is the best diagnostic to distinguish between SFN and AGN galaxies,
whereas [OI]- and [SII]- diagrams are less efficient. They also
claimed that it is usually possible to distinguish between SFN and AGN
using the [NII]/H$\alpha$ ratio only.

\begin{figure}
\epsscale{0.5}
\plotone{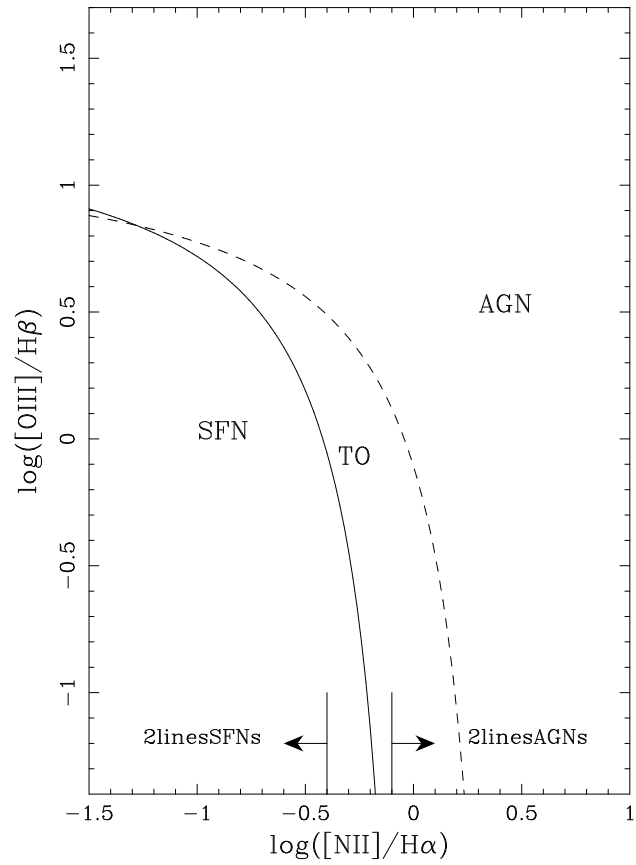}
\caption{Diagnostic diagram with classification criteria used in this
study. Solid line corresponds to Ka03 sequence and the dashed line to
the Ke01 sequence. Vertical lines corresponds to values for the
[NII]/H$\alpha$ ratio of -0.4 y -0.1 as explained in the
text. \label{fig4}}
\end{figure}

\begin{figure*}
\epsscale{1}
\plotone{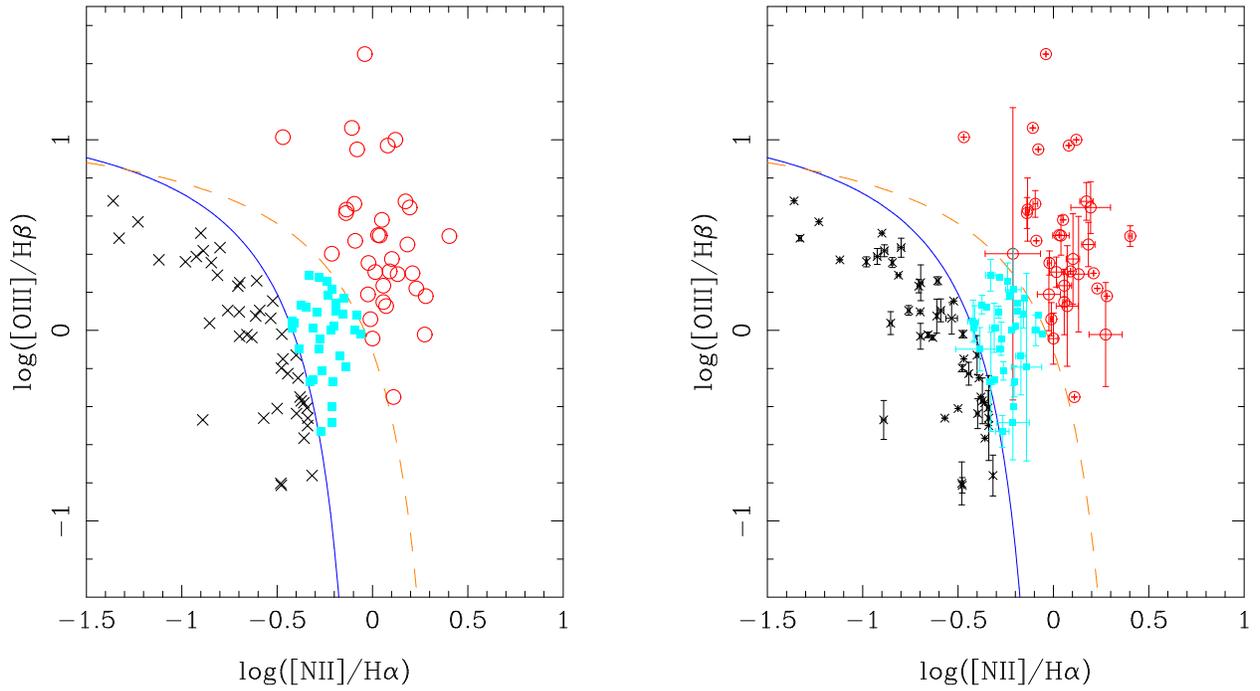}
\caption{BPT diagram for the galaxies having the four main emission
lines. Solid line was defined by \citet{kau03} and surround
SFNs. Dashed line corresponds to the extreme starburst classification
line from \citet{kew01}. Crosses are SFNs, open circles are AGNs and
full squares correspond to TOs.On the right plot error bars are
plotted over each galaxy.See the electronic edition of the Journal for
a color version of this figure. \label{fig5}}
\end{figure*}

We will base our classification on the criteria suggested by
\citet{kew06} and \citet{sta06}. In Fig.~\ref{fig4} we present
an empty BPT diagram to illustrate the classification criteria we have
adopted in this study. The two curves correspond respectively to the
Ka03 (solid line) and the Ke01 (dashed line) limits. Using both
sequences, we have classified the galaxies as pure AGNs if they are
located above the Ke01 sequence and as SFNs if they lie below the Ka03
sequence. Galaxies located between the two sequences were consequently
classified as TOs. The two vertical lines in the bottom part of the
diagram correspond to values for log([NII]/H$\alpha$) of -0.4 and -0.1
respectively. We have used these limits to classify those galaxies for
which we have confident values only for the [NII]/H$\alpha$ ratio. We
have classified as SFNs those galaxies with log([NII]/H$\alpha$)
$\leqslant$-0.4 and as pure AGNs those with log([NII]/H$\alpha$) $ >$
-0.1 (corresponding to a [NII]/H$_{\alpha}$ ratio of 0.8).  Again,
galaxies with line ratios in between were classified as TOs. The
adopted value for the SFNs (-0.4) is the same as in
\citet{sta06}. However, we have restricted a bit more the AGN limit
(-0.1 instead of their -0.2). This appears more realistic if we take
into account the observed value of this ratio for our TOs with all
four lines in the BPT diagram (see Fig~\ref{fig5}).

Although we did not impose any restriction to the S/N on the lines,
all the identified emission galaxies have lines detected at more than
3$\sigma$ confidence level.

%
\subsection{Nuclear Activity for the HCG sample}
\label{sec-5.2}
%

The majority (111 or 66\%) of the 169 emission line galaxies in our
sample, have accurate measurements at least for the four most
important emission lines: H$\beta$, [OIII], H$\alpha$ and [NII]. For
21 galaxies (12\% of the emission galaxies) we obtained valid
measurements for [OIII], H$\alpha$ and [NII] lines, and in many cases
also for the [SII] lines, but H$\beta$ emission is missing due mainly
to the weakness of the line compared to the strong stellar
absorption. In 33 galaxies (19\%) we only acquired measurements for
the red lines, in particular for the [NII] and H$\alpha$ lines, due to
a relatively low S/N in the blue range of their spectrum or because
they were observed only with the red grism, so they can not be located
in the BPT diagram.  For the remaining galaxies (3\%) we could not
calculate any line ratio: in three galaxies the most conspicuous line
corresponds to H$\alpha$ and for one, from the literature, only the
[NII]$\lambda$6583\AA\ line was detected.

The assigned nuclear activity classification for each galaxy is listed
in the last column of tables \ref{tab4} and \ref{tab5}
respectively for the new observed galaxies and for those coming from
the archives and literature. Emission galaxies have been classified as
SFNs, AGNs, and TOs according to the criteria described in the
previous section.

The 111 galaxies having four line measured are represented on the BPT
diagram in Figure \ref{fig5}, using crosses for the SFNs, open
circles for pure AGNs and filled squares for the TOs. In the figure on
the right is showed the location of each galaxy with the error bars
determined, as estimated in section \ref{sec-3.3}.  According to this
diagram, 34 galaxies are AGNs, 33 are TOs and 44 SFNs.

\begin{figure}
\epsscale{0.5}
\plotone{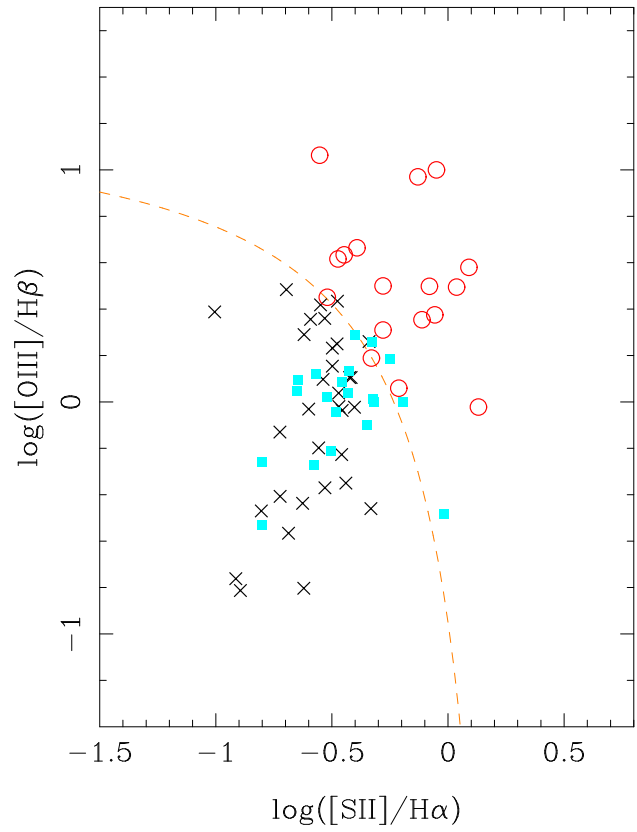}
\caption{[SII]-diagram for the 70 galaxies from the BPT with also [SII]
line measurements. Dashed line corresponds to the extreme starburst
classification line from \citet{kew01}. Symbols as in
Figure~\ref{fig5}.See the electronic edition of the Journal for a
color version of this figure. \label{fig6}}
\end{figure}

Although our classification is based mainly on the BPT diagram, we
have also used the [SII]- and [OI]- diagrams, in the case of reliable
detection. 70 of the 111 galaxies from the BPT diagram have
measurements for the [SII] lines.  Figure \ref{fig6} shows the
location of these galaxies in the [SII]-diagram, where the symbols
correspond to the activity type as determined by the BPT diagram. In
this figure the dashed line represents the Ke01 sequence, which mainly
separates AGNs from the other activity types. As can be seen from the
comparison of both figures, there is good concordance between
the diagrams. Basically both AGNs and SFNs in the BPT diagram maintain
their classification in the [SII]-diagram.  We have reclassified 4 TOs
from the [NII]-diagram into TO/AGNs taking into account the derived
[SII]/H$\alpha$ ratios, and the general behavior of their
spectra. Their [SII]/H$\alpha$ values place them above the Ke01
sequence in the [SII]-diagram, suggesting that the AGN component is
the dominant feature.

\begin{figure}
\epsscale{0.5}
\plotone{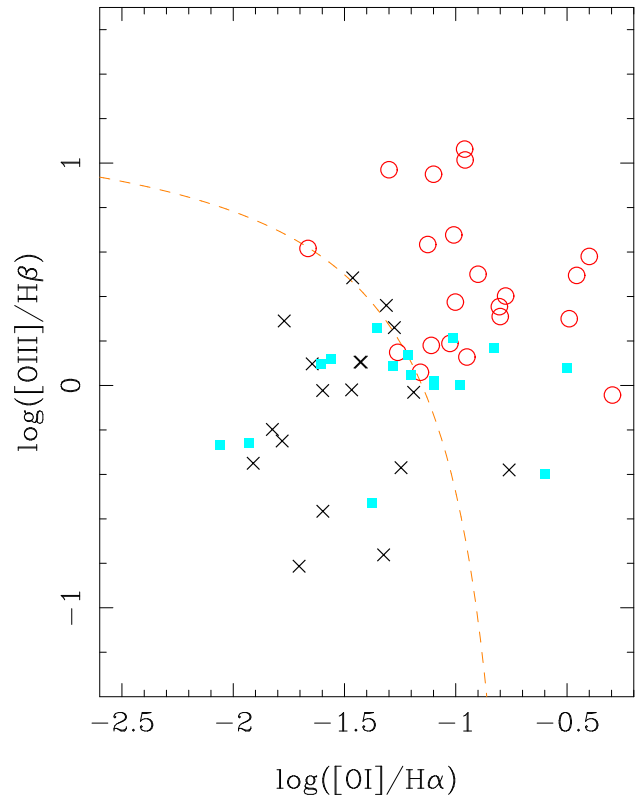}
\caption{[OI]-diagram for the 55 galaxies with measurement of [OI]
line. Dashed line corresponds to the extreme star formation line
defined by \citet{kew01} for this diagram. See Figure~\ref{fig5} for
symbols.See the electronic edition of the Journal for a color version 
of this figure. \label{fig7}}
\end{figure}

Figure~\ref{fig7} shows the [OI]-diagram for 55 of the 111
galaxies from the BPT diagram with value of the [OI] line. The symbols
again correspond to the activity type as determined based on the BPT
diagram. The dashed curve represents the corresponding Ke01 sequence
in this diagnostic diagram. As in the previous [SII]-diagram, the
majority of the galaxies maintain their BPT classification in the
[OI]-diagram. We have reclassified 7 TOs from the [NII]-diagram as
TO/AGN, taking into account the [OI]/H$\alpha$ ratios, which place
them well into the AGN region. One of these galaxies was already
classified as TO/AGN using the [SII]-diagram. Hereafter, we will count
the 10 TO/AGNs as AGNs.

Summarizing our classification, for 111 galaxies with the four main
emission lines measured we find 44 SFNs, 23 TOs and 44 AGNs.

For galaxies classified as AGNs we can further distinguish between Sy2
and LINER.  Using the same criterion as in \citet{coz98} to separate
low from high excitation AGNs, we count about 37\% of Sy2. We found
the same proportion using the sequence proposed by
\citet{kau03}. Therefore our data indicate that 37\% of the AGN
population in the sample host Seyfert activity type.

For the 21 emission galaxies with values for [OIII], H$\alpha$ and
[NII] lines, but without direct measurement of the H$\beta$ line, and
for the 33 galaxies without data for the [OIII] and H$\beta$ lines, we
have based our classification on the [NII]/H$\alpha$ ratio alone, as
explained in section \ref{sec-5.1}. We find 8 SFNs, 16 TOs and 30
AGNs.

For the remaining 4 galaxies, one (H21c) comes from the literature and
the other three (H52b, H76f and H95d) have been observed by us.
\citet{coz00} classified H21c as LLAGN, with only [NII]$\lambda$6583
detected. H76f and H95d show only H$\alpha$ as the more conspicuous
line over the continuum and we classify them as SFNs. In H52b we
detect four emission lines: [NII]$\lambda$6548\AA, H$\alpha$ and the
two [SII] lines while [NII]$\lambda$6583\AA\ is inside the atmospheric
band, due to the redshift of the galaxy. To obtain its nuclear
classification we have estimated the [NII]$\lambda$6583\AA\ flux as 3
times the measured [NII]$\lambda$6548\AA\ intensity, obtaining a
[NII]/H$\alpha$ ratio of 1.56 which corresponds to an AGN.

Summarizing our results, 63\% of the galaxies (169 of 270) in our HCG
sample shows emission lines. Using different diagnostic diagrams, we
classify 54 as SFNs, 39 as TOs and 76 as AGNs. These represent
respectively 32\%, 23\% and 45\% of the emission line galaxies. This
confirm that AGN is the most frequent activity type in HCGs. For the
whole sample, 37\% show no emission, 20\% are SFN, 15\% are TOs, 28\%
are AGNs.

%
\subsection{AGN luminosity}
%

\begin{figure}
\epsscale{1}
\plotone{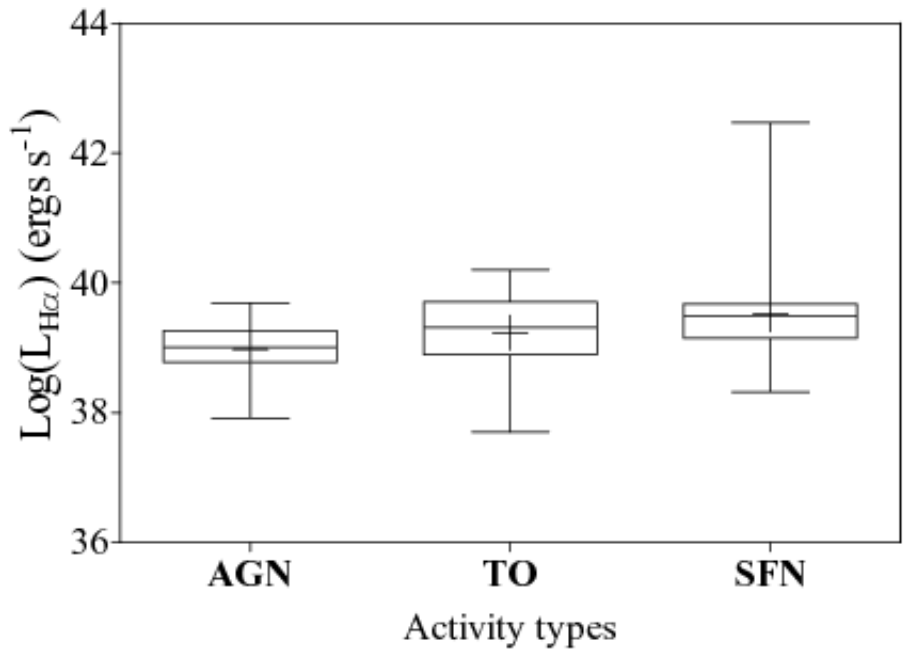}
\caption{Box-whiskers plot for the observed nuclear H$\alpha$ luminosity
of the galaxies with different activity type. The upper and lower
limits of the boxes are the 75\% and 25\% percentiles
respectively. The extent of the vertical bars indicate the full range
of the data (from minimum to maximum). \label{fig8}}
\end{figure}

We have estimated the observed nuclear luminosity of the emission line
galaxies in our sample by using their measured H$\alpha$ fluxes
together with their corresponding redshifts. From the 200 new
observations, 122 have emission lines and from them 118 have absolute
flux calibration. From this subsample, 39 galaxies are classified as
SFNs, 26 as TOs and 53 as AGNs. In Figure~\ref{fig8} we show the
box-whisker plot for the distribution of luminosity separated by
activity types. The SFNs with a median log(L$_{H\alpha}$) of 39.5 seem
to be slightly more luminous than the AGNs, log(L$_{H\alpha}$) =39.0,
while the TOs with a median value of 39.3 look intermediate between
the two. A non parametric statistical test (Kruskall-Wallis) confirms
an extremely significant difference (P $< 0.0001$) between the AGNs
and SFNs at a 95\% confidence level. No significant statistical
differences are accounted from the test between the AGNs and TOs or
SFNs and TOs.

In \citet{coz98,coz04}, and \citet{mar08a,mar08b}, it was pointed out
that AGNs in CGs are mostly LLAGNs. Our more complete sample allows us
to confirm this claims for the HCGs. The observed H$\alpha$ luminosity
for the 53 AGNs ranges from 8.3$\times$10$^{37}$ to
4.9$\times$10$^{39}$erg\ s$^{-1}$, with a median of
1.0$\times$10$^{39}$erg\ s$^{-1}$. These values are below the upper
limits of 10$^{40}$-10$^{41}$ erg\ s$^{-1}$ \citep{ho97,zha07} which
are used to define LLAGNs. This is also verified if we include the 26
TOs.  For the resulting 79 galaxies, the observed H$\alpha$ luminosity
ranges from 5.0$\times$10$^{37}$ to 1.58$\times$10$^{40}$erg\ s$^{-1}$
with a median value of 1.1$\times$10$^{39}$erg\ s$^{-1}$, confirming
that they also would correspond to Low Luminosity AGNs.

For all galaxies hosting an AGN, 43 galaxies have direct measurement
of H$\beta$. For these galaxies we have estimated the extinction
corrected H$\alpha$ luminosity. Using the extinction law from
\citet{car89} and an intrinsic relation H$\alpha$/H$\beta$=3.1, to
correct from internal extinction, the corrected L$_{H\alpha}$ ranges
from 7.4$\times$10$^{37}$ to 9.0$\times$10$^{40}$erg\ s$^{-1}$, with a
median value of 7.1$\times$10$^{39}$erg\ s$^{-1}$. Even after
extinction correction the AGNs are still below the upper limits
suggested for the LLAGNs.

In our whole sample, we have found only one genuine Sy1, H91a.
\citet{mar08a} showed, using two different samples of CGs, that there
is a significant deficiency of broad-line AGNs compared to other
environments. This results suggests that in CGs AGNs have broad line
regions which are too small to be detected, or, more simply, they do
not have such regions. Our present results show that
not only BLAGNs are deficient in HCGs but so are high luminosity AGNs
in general.

%
\section{Summary and Conclusions}
%

In this paper we presented the nuclear spectral classification for 270
galaxies in 64 HCGs. For this classification we have obtained new
spectroscopic observations for 200 member-galaxies, added 11 spectra
from ESO and 6dF Archives and for 59 galaxies collected emission line
ratios from the literature. After template subtraction we found that a
large fraction of these galaxies, 63\%, show emission lines,
confirming previous spectroscopic results based on smaller samples:
62\% of emission for 91 galaxies in 27 HCGs \citep{coz98,coz04} or
70\% on a limited number of the 63 brightest galaxy members in 28 HCGs
\citep{shi00}. Comparable fractions of emission line galaxies was also
found in other samples of compact groups: 72\% in the South Compact
Group Catalogue \citep{coz00} and 68\% in Compact Groups from the
Update Zwicky Catalogue \citep{mar07,mar08c}. In general, therefore,
CGs are not deficient in emission-line galaxies.

Using standard diagnostic diagrams we have classified 45\% of the
emission-line galaxies as AGNs, 32\% as SFNs and 23\% as TOs. This
makes AGN the most frequent activity type encountered in HCG emission
galaxies. As we mentioned in Section\ref{sec-5.1}, recent
observations of TOs show the presence of radio and X-ray compact cores
\citep{fil02,fil04,nag05,ho08} and clear indications that the majority
of TOs harbor an AGN \citep[see e.g. review by][and references
therein]{ho08}. Thus, considering that TOs also host AGNs in their
nuclei, the fraction of AGNs in HCGs emission galaxies rise to 68\%
and for the whole sample AGNs form 43\%, non-emission galaxies form
37\% and SFNs form only 20\%.

To realize how important AGNs are in HCGs, one has to compare with
what is found in other environments. To do this comparison we consider
the fraction of galaxies hosting an AGNs (pure AGNs plus TOs) over the
emission galaxy population found in other studies. One
important spectroscopic survey of the general field in the nearby
Universe is the Palomar sample, with galaxies brighter than 12.5mag in
B and mainly located in loose groups, low-density environment and
Virgo cluster. For this survey \citep{ho97} reports rates of 50\% AGNs
(35\% AGNs + 15\% TOs) and 50\% SFNs. Similar results were found by
\citet{cart01} based on the 15R-North galaxy redshift
survey. \citet{kau03} using a wide sample of narrow-line AGN from the
SDSS and using classification criterions comparable to ours, obtain
40\% AGNs compared to 60\% SFNs. \citet{mil03} reported 26\% to 40\%
AGNs based on a similarly selected sample. The AGN ratio decrease even
more if we consider isolated galaxies; \citet{sab08,sab09} found that
the fraction of FIR and optical selected AGN-candidates over a sample
of isolated galaxies range between 7\% and 22\%.

Indeed the AGN fraction (68\%) we found in the HCGs is therefore
significantly higher than what is found in the general field or in
isolated galaxies. When going to galaxy clusters, which constitute the
opposite end of environment, the difference is even greater. 
\citet{dre85,dre99} evaluated the fraction of AGNs to be only
1\% over the whole sample, percentage that increases to 5-8\% when
X-ray data are also included \citep{mart07}. AGNs seem to avoid galaxy
clusters. However these percentages should be taken with caution due
the lack of enough comparable optical data available for cluster
galaxies.

But what happens if one constrains further the environment?. For two
different samples of close galaxy pairs, \citet{woo07} and
\citet{foc08} report that the fraction of AGNs can reach 40\% of the
emission galaxies. This rate, thought enhanced respect to isolated
galaxies, is similar to that found in the field and clearly lower than
what we detect in compact groups.  Therefore AGNs inhabit more
frequently in HCGs than in close pairs. Even if galaxy-galaxy
encounters also occur in pairs of galaxies, this only mechanism could
not be so efficient to feed the AGN activity in pairs as in compact
group environment, suggesting that direct galaxy-galaxy interaction
alone will be not enough to explain the AGN fraction found in HCGs and
that other parameters, like for example the evolutionary state of the
structure must be considered.

A high number of AGNs seem, therefore, like a proper trait of the
HCGs. However, another important trait seems to be the low luminosity
of these AGNs. After extinction correction we found a median AGN
H$\alpha$ luminosity of L$_{H\alpha}$=7.1$\times$10$^{39}$ erg
s$^{-1}$, which confirms that AGNs in HCGs correspond mainly to Low
Luminosity AGNs \citep{coz04,mar08b,mar08c}. From the 76 galaxies
classified as AGN, only two (H5a and H91a) show broad components: H5a
is a Sy1.9 (almost a Sy 2) and H91a a Sy 1.2. This gives a ratio of
broad to narrow line AGNs of only 3\%, which makes HCGs deficient in
such AGNs. This is in agreement with the deficit of BLAGNs found
\citet{mar08a} based in a smaller sample of HCGs. An extensive
discussion about biases and detection limits can be found in that
work.

The low percentage of galaxies hosting SFNs, only 20\% of the total
sample and 32\% of the emission galaxies, is also quite
remarkable. It clearly indicates that there is no enhancement of star
formation in HCGs \citep{igl99,coz00}. This result is in good
agreement with the truncation of star formation previously detected in
early-type HCG galaxies \citep{ros07}, the nearly complete suppression
of star formation reported in groups like H79 and H92
\citep{dur08,sul01} and the relatively low FIR emission observed in
HCGs in general \citep{verd98}. Also, spiral galaxies in HCGs seem
particularly deficient in HI, with a mean value of only 24\% of the
expected gas content observed \citep{verd01}. So, in terms of
activity, the main characteristic of HCGs seems related to a general
absence or severe deficiency of gas. The low luminosity of the AGNs
and absence of Broad Line Regions could be a direct consequence of
this characteristic. The few gas remaining in the nuclei is not enough
for broad line regions to form and the resulting low accretion rates
originate a low luminosity AGN.

\acknowledgements
MAM acknowledges Ministerio de Educacion y Ciencia for financial
support grant FPU AP2003-4064. MAM, AdO and JP are partially supported
by spanish research projects AYA 2006-14056, AYA 2007-62190,
P08-TIC-03531 and TIC114. We thank the referee for constructive
comments. We thank the TAC of the Observatorio Astron\'omico Nacional
at San Pedro M\'artir for time allocations. We acknowledge also the
usage of the Hyperleda database (http://leda.univ-lyon1.fr). This
research has made use of the NASA/IPAC Extragalactic Database (NED)
which is operated by the Jet Propulsion Laboratory, California
Institute of Technology, under contract with the National Aeronautics
and Space Administration. We acknowledge the entire 6dFGS team
(www.aao.gov.au/local/www/6df/6dFGSteam.html) and the use of ESO
archive data.\\

%

\clearpage


\begin{deluxetable}{lllcccrlllccr}
\tabletypesize{\footnotesize}
\tablecolumns{13}
\tablewidth{0pt}
\tablecaption{Complete sample of HCGs\label{tab1}}
\tablehead{
\colhead{Name} & \colhead{$\alpha$} & \colhead{$\delta$} & \colhead{N$_{g}$} &\colhead{N$_{s}$} & \colhead{V$_r$} &\vline& \colhead{Name} & \colhead{$\alpha$} & \colhead{$\delta$} & \colhead{N$_{g}$} &\colhead{N$_{s}$}& \colhead{V$_r$} \\
       & \colhead{J2000}    &  \colhead{J2000}   &        &      &\colhead{(km/s)}&\vline && \colhead{J2000}    &  \colhead{J2000}   &        &      &\colhead{(km/s)}}
\startdata
H1     & 00 26 00.2 & +25 43 05 & 4  & 4   &  10158  &\vline &H56    & 11 32 39.6 & +52 56 25 & 5  & 5   &   8116  \\	
H3     & 00 34 27.4 & -07 35 35 & 4  & 4   &   7835  &\vline &H57    & 11 37 50.5 & +21 59 06 & 8  & 8   &   9087  \\	
H4     & 00 34 15.9 & -21 26 48 & 3  & 3   &   7770  &\vline &H59    & 11 48 26.6 & +12 42 40 & 4  & 4   &   4113  \\	
H5     & 00 38 54.3 & +07 03 36 & 4  & 4   &  12405  &\vline &H61    & 12 12 24.9 & +29 11 21 & 3  & 3   &   3919  \\	
H6     & 00 39 10.1 & -08 23 43 & 4  & 4   &  11118  &\vline &H62    & 12 53 08.1 & -09 13 27 & 4  & 4   &   4148  \\	
H7     & 00 39 23.9 & +00 52 41 & 4  & 4   &   4243  &\vline &H63    & 13 02 10.0 & -32 46 05 & 3  & 0   &   9316  \\	
H10    & 01 26 13.0 & +34 42 41 & 4  & 4   &   4794  &\vline &H64    & 13 25 43.3 & -03 51 28 & 3  & 3   &  10699  \\	
H13    & 01 32 22.1 & -07 52 52 & 5  & 4   &  12436  &\vline &H67    & 13 49 03.5 & -07 12 20 & 4  & 4   &   7385  \\
\enddata
\tablecomments{Group radial velocities have been calculated using galaxy radial velocity extracted from our spectra, otherwise, radial velocity was taken from HYPERLEDA database.}
\tabletypesize{\normalsize}
\end{deluxetable}

\begin{deluxetable}{llllllllll}
\tabletypesize{\normalsize}
\tablecolumns{10}
\tablewidth{0pt}
\tablecaption{Instrumental Configurations\label{tab2}}
\tablehead{
\colhead{Obs.}&  \colhead{Tel.}  &\colhead{Spectrograph}& \colhead{Detector} &\colhead{pixel}& \colhead{Dimension} & \colhead{Scale} & \colhead{Grism}&  \colhead{Range}& \colhead{Disp.} \\ 
    & & & & \colhead{$\mu$m/px} && \colhead{``/px} & & \colhead{\AA} &
    \colhead{\AA/px}}
\startdata
CAHA  &  2.2m    & CAFOS     &SITE        &24 &2048$\times$2048  &  0.53  &B100     &   3600-6200    &  2    \\
      &          &           &            &   &                  &        &G100     &   4300-7800    &  2.12 \\
ORM   &  2.5m    & ALFOSC    &E2V         &15 &2048$\times$2048  &  0.19  &GR4      &   3600-8000    &  3    \\
      & (NOT)    &           &            &   &                  &        &GR8      &   5825-8350    &  1.24 \\
OSN   &  1.5m    & ALBIREO   &Loral/Leser &15 &2048$\times$2048  &  0.90  &Red4     &   3600-7500    &  2    \\
SPM   &  2.1m    & Boller-Chivens &SITE   &24 &1024$\times$1024  & 1.05   &R300     &   3800-7500    &  4.3  \\ 
\enddata
\tabletypesize{\normalsize}
\end{deluxetable}

\begin{deluxetable}{lccrccccc}
\tabletypesize{\normalsize}
\tablecolumns{9}
\tablewidth{0pt}
\tablecaption{Log of the Observations\label{tab3}}
\tablehead{
\colhead{Id.} &  \colhead{Date} & \colhead{Conf} & \colhead{PA} & \colhead{T$_{exp}$} & \colhead{Aimass} & \colhead{Slitwidth} & \colhead{Ap.} & \colhead{Emission}\\
   &    &    & \colhead{($^o$)} & \colhead{(s)}   &      & \colhead{(``)}& \colhead{(kpc)} & }
\startdata
  H1a  &    18 Nov 2004   & B100+G100    & 36.2    & 2400+2400  & 1.04 & 1.5 & 1.13 & Yes \\ 
  H1b  &    18 Nov 2004   & B100+G100    & 36.2    & 2400+2400  & 1.04 & 1.5 & 1.13 & No  \\ 
  H1c  &    19 Nov 2004   & B100+G100    & 84.5    & 3200+3400  & 1.02 & 2.0 & 1.11 & Yes \\ 
  H1d  &    19 Nov 2004   & B100+G100    & 84.5    & 3200+3400  & 1.02 & 2.0 & 1.11 & Yes \\ 
  H3a  &    01 Nov 2005   & GR4          & 148.3   & 2400       & 1.30 & 1.0 & 1.00 & Yes \\ 
  H3b  &    01 Nov 2005   & GR4          & 148.3   & 2400       & 1.30 & 1.0 & 1.00 & Yes \\ 
  H3c  &    01 Nov 2005   & GR4          & 84.1    & 3000       & 1.24 & 1.0 & 1.00 & Yes \\ 
  H3d  &    01 Nov 2005   & GR4          & 84.1    & 3000       & 1.24 & 1.0 & 1.00 & No  \\
\enddata
\tabletypesize{\normalsize}
\tablenotetext{*}{Galaxies with discordan redshift that do not belong to the final sample.}
\end{deluxetable}

\begin{deluxetable}{lccccrrr}
\tabletypesize{\normalsize}
\tablecolumns{8}
\tablewidth{0pt}
\tablecaption{Line ratios of the observed galaxies\label{tab4}}
\tablehead{
\colhead{Gal.} &  \colhead{log$\frac{[OIII]}{H\beta}$} & \colhead{log$\frac{[OI]}{H\alpha}$} & \colhead{log$\frac{[NII]}{H\alpha}$} & \colhead{log$\frac{[SII]}{H\alpha}$}  & \colhead{F(H$\alpha$)} & \colhead{L(H$\alpha$)} & \colhead{Cla}}
\startdata
 H01a &   -0.020$\pm$0.019 &   -1.468$\pm$0.045 &   -0.475$\pm$0.004 &   -0.470$\pm$0.045& 0.46E-14 &  40.07  &    SFN \\ 
 H01c &    0.498$\pm$0.103 &           -        &    0.039$\pm$0.043 &   -0.080$\pm$0.065& 0.11E-14 &  39.44  &    AGN \\ 
 H01d &           -        &           -        &    0.133$\pm$0.052 &   -0.378$\pm$0.072& 0.36E-15 &  38.95  &    AGN \\ 
 H03a &    0.677$\pm$0.099 &   -1.009$\pm$0.071 &    0.173$\pm$0.035 &           -       & 0.49E-15 &  38.80  &    AGN \\ 
 H03b &   -0.193$\pm$0.493 &           -        &   -0.141$\pm$0.079 &           -       & 0.18E-14 &  39.43  &     TO \\ 
 H03c &   -0.813$\pm$0.040 &   -1.704$\pm$0.048 &   -0.478$\pm$0.007 &   -0.894$\pm$0.048& 0.15e-14 &  39.31  &    SFN \\
 H05a &    0.495$\pm$0.054 &   -0.457$\pm$0.049 &    0.403$\pm$0.009 &    0.037$\pm$0.049& 0.14E-14 &  39.69  &    AGN \\ 
 H05b &           -        &           -        &    0.410$\pm$0.192 &           -       & 0.22E-16 &  37.91  &    AGN \\
\enddata
\tabletypesize{\normalsize}
\tablenotetext{*}{H26a and H40c have no error because they have been observed with GR8 where not templates were availables. We give line flux and luminosities when spectra have been absolute flux calibrated. Values with two colons indicates that errors in fotometry are greater than 20\%.}
\end{deluxetable}

\begin{deluxetable}{lccccrclccccr}
\tabletypesize{\footnotesize}
\tablecolumns{13}
\tablewidth{0pt}
\tablecaption{Line ratios of the archive and literature data\label{tab5}} 
\tablehead{
\colhead{Gal.} &  \colhead{log$\frac{[OIII]}{H\beta}$} & \colhead{log$\frac{[OI]}{H\alpha}$} &\colhead{log$\frac{[NII]}{H\alpha}$}& \colhead{log$\frac{[SII]}{H\beta}$ }  & \colhead{Cla}& \vline& \colhead{Gal.} &  \colhead{log$\frac{[OIII]}{H\beta}$} & \colhead{log$\frac{[OI]}{H\alpha}$} &\colhead{log$\frac{[NII]}{H\alpha}$}& \colhead{log$\frac{[SII]}{H\beta}$ }  & \colhead{Cla}}
\startdata
 H04a    &  -0.260   &   -1.930   &   -0.310  &  -0.800   &   TO  & \vline &  H56b  &    1.450   &      -     &   -0.040  &     -     &    AGN  \\
 H04b*   &     -     &   -0.911   &   -0.199  &      -    &   TO  & \vline &  H56d  &   -0.100   &      -     &   -0.280  &     -     &     TO  \\
 H04d*   &  -0.566   &   -1.597   &   -0.359  &  -0.687   &  SFN  & \vline &  H56e  &      -     &      -     &   -0.480  &     -     &    SFN  \\
 H16a    &   0.310   &   -0.800   &    0.090  &  -0.280   &  AGN  & \vline &  H62a  &    0.000   &   -1.100   &   -0.220  &  -0.320   & TO/AGN  \\
 H16b    &   0.580   &   -0.400   &    0.050  &   0.090   &  AGN  & \vline &  H67b  &   -0.150   &      -     &   -0.470  &     -     &    SFN  \\
 H16c    &   0.120   &   -1.560   &   -0.350  &  -0.570   &   TO  & \vline &  H68b  &    0.140   &      -     &   -0.190  &     -     &     TO  \\
 H16d    &  -0.350   &   -1.910   &   -0.380  &  -0.440   &  SFN  & \vline &  H68c  &   -0.020   &      -     &   -0.060  &     -     &     TO  \\
 H21a*   &     -     &      -     &   -0.421  &     -     &  SFN  & \vline &  H79a  &    0.280   &      -     &   -0.280  &     -     &     TO  \\
\enddata
\tabletypesize{\normalsize}
\tablenotetext{*}{These galaxies correspond to the Archive spectra. The rest of the galaxies come from the literature. There is no line ratio available for H21c, we have taken its nuclear classification from \citet{coz00}}
\end{deluxetable}

\end{document}